# Establishing a non-hydrostatic global atmospheric modeling system (iAMAS) at 3-km horizontal resolution with online integrated aerosol feedbacks on the Sunway supercomputer of China

Jun Gu[1#], Jiawang Feng[1#], Xiaoyu Hao[2#], Tao Fang[2#], Chun Zhao[1,3,4*], Hong An[2*], Junshi Chen[2], Mingyue Xu[1], Jian Li[5], Wenting Han[2], Chao Yang[6], Fang Li[7], Dexun Chen[8]

[1]School of Earth and Space Sciences, University of Science and Technology of China, Hefei 230026, China

[2]School of Computer Science and Technology, University of Science and Technology of China, Hefei 230026, China

[3]CAS Center for Excellence in Comparative Planetology, University of Science and Technology of China, Hefei, China

[4]Frontiers Science Center for Planetary Exploration and Emerging Technologies, University of Science and Technology of China, Hefei, China

[5]Chinese Academy of Meteorological Sciences, Beijing 100081, China

[6]School of Mathematical Sciences, Peking University, Beijing 100871, China

[7]Jiangnan Institute of Computing Technology, Wuxi, Jiangsu, China

[8]Tsinghua University, Beijing, China

[#]These authors contributed equally: Jun Gu, Jiawang Feng, Xiaoyu Hao, Tao Fang

* Corresponding authors: Chun Zhao (chunzhao@ustc.edu.cn) and Hong An (han@ustc.edu.cn)

# Abstract

During the era of global warming and highly urbanized development, extreme and high impact weather as well as air pollution incidents influence everyday life and might even cause the incalculable loss of life and property. Although with the vast development of numerical simulation of atmosphere, there still exists substantial forecast biases objectively. To predict extreme weather, severe air pollution, and abrupt climate change accurately, the numerical atmospheric model requires not only to simulate meteorology and atmospheric compositions and their impacts simultaneously involving many sophisticated physical and chemical processes but also at high spatiotemporal resolution. Global atmospheric simulation of meteorology and atmospheric compositions simultaneously at spatial resolutions of a few kilometers remains challenging due to its intensive computational and input/output (I/O) requirement. Through multi-dimension-parallelism structuring, aggressive and finer-grained optimizing, manual vectorizing, and parallelized I/O fragmenting, an integrated Atmospheric Model Across Scales (iAMAS) was established on the new Sunway supercomputer platform to significantly increase the computational efficiency and reduce the I/O cost. The global 3-km atmospheric simulation for meteorology with online integrated aerosol feedbacks with iAMAS was scaled to 39,000,000 processor cores and achieved the speed of 0.82 simulation day per hour (SDPH) with routine I/O, which enables us to perform 5-day global weather forecast at 3-km horizontal resolution with online natural aerosol impacts. The results demonstrate the promising future that the increasing of spatial resolution to a few kilometers with online integrated aerosol impacts may significantly improve the global weather forecast.

## 1. Introduction

Numerical prediction of weather, air quality, and climate with atmospheric model supports our daily lives and even influences national policy and strategy development. Since the first atmospheric model that was written on ENIAC (often considered as the first computer in history) to today [1], the atmospheric model has evolved from a simple solver of a few equations to one of the world's most complicated software (such as the Weather Research and Forecasting (WRF) model [2], Community Atmospheric Model (CAM) [3], Model for Prediction Across Scale for Atmosphere (MPAS-Atmosphere) [4], Nonhydrostatic ICosahedral Atmospheric Model (NICAM) [5], Finite-Volume Module of Integrated Forecasting System (IFS-FVM) [6], etc.) that possesses millions of lines of code, and involves hundreds or even thousands of contributors from various research institutions and organizations. Nowadays, atmospheric model is an important tool for simulating past, present, and future weather, climate, and air quality. It is also an important component of the Earth System Model (ESM) for understanding the earth's water and energy cycle.

Atmospheric model for more accurate forecast has been improved for last few decades. However, there is still a large gap between the scales of the atmospheric phenomena that need to be resolved and the spatial resolution that computers can afford, because the global atmospheric domain is extremely large (Fig. 1a). With coarse horizontal resolution, the sub-grid process such as deep convection and orographic forcing are often parameterized as a scheme in the model. Previous studies have pointed out that these parameterizations caused significant uncertainties for weather and climate predictions [7]. In the last decade, with the increase of computing resources, the global non-hydrostatic atmospheric models, that are suitable for global simulations at the horizontal resolution of a few kilometers, have been developed. Many studies recognized that global atmospheric simulations at the horizontal resolutions of several kilometers would improve forecasting accuracy. Nevertheless, atmospheric simulation at horizontal resolutions of several kilometers requires significantly more computational cost compared to traditional ones at tens or hundreds of kilometers. For global atmospheric models, at least an increase of $2^3$ computational workload is required with the doubling of horizontal resolution.

Besides the requirement of high spatiotemporal resolutions, the inclusion of atmospheric compositions such as aerosols and their impacts on meteorology through scattering and absorbing radiation and indirectly by modifying microphysical properties of clouds was also recognized as a critical factor for weather and climate prediction (Fig. 1b), particularly over the regions with high loading of aerosols [8-12]. In addition, appropriate simulation of atmospheric composition is also required to predict air quality. However, modeling appropriately atmospheric compositions and their impacts simultaneously with meteorology requires many additional sophisticated processes, which can significantly increase the computational cost, slow down the simulation efficiency, and add I/O burden. One reason for the significant increase of computational and I/O cost of including atmospheric compositions is that much more (hundreds) prognostic chemical species need to be simulated compared to meteorological fields (ten to twenty). Therefore, due to the limit of computational resource and efficiency, so far most weather and climate simulations at convection-permitting horizontal resolutions (several kilometers) neglected or included prescribed chemical compositions [13-20].

Several studies have been conducted to explore the technical/scientific issues that are critical for global atmospheric simulations at the kilometer scale with the powerful High-Performance Computing (HPC). Many of them targeted to only assess the performance of the dynamic cores of atmospheric models [2, 15, 21-23]. For example, some institutions have already demonstrated the capability of global simulations with 3-km mesh using two state-of-the-art atmospheric dynamic cores (MPAS and FV3) [21]. The global simulations with different dynamic cores with a mesh of less than 10-km were inter-compared in the DYnamics of the Atmospheric general circulation Modeled On Non-hydrostatic Domains (DYAMOND) project [23]. The fully implicit atmospheric dynamical core achieved an idealized experiment with a 488-m mesh (1.1 SDPH) [22]. Some studies started real-world global simulations with several kilometers mesh but only with hydrostatic dynamic core to increase computational efficiency (through using larger integration time step) compared to non-hydrostatic dynamic core [17, 24]. However atmospheric simulations with a sub-10km mesh using hydrostatic dynamic core may introduce significant errors and are not desired [25]. Among those previous studies, only a few achieved real-world non-hydrostatic global atmospheric simulations with horizontal resolutions of several kilometers. In addition, some coupled global high-resolution atmospheric model with aerosol transport model [26, 27] through "offline" or "online access" methods[28]. They conducted global simulations at the horizontal resolutions of a few kilometers to assess the resolution impact on aerosol transport and aerosol feedback on clouds [26, 27]. So far, few studies reported the effort to conduct the global non-hydrostatic atmospheric simulation at a few kilometers with meteorology and aerosol online integrated (i.e., simulate meteorology and aerosol and their interactions simultaneously in one model and with one main time step for integration) that may provide more accurate modeling results but is often more computationally expensive [28, 29]. In particular, few of them assessed the computational performance of meteorology-aerosol online integrated model with routine and frequent I/O.

Moreover, with the leading-edge supercomputer with heterogeneous architecture, some studies were providing evidence that the heterogeneous architecture involving many-core accelerators may improve the computational efficiency and is likely becoming the mainstream for modern supercomputer systems. The Sunway TaihuLight in China is a typical supercomputer with heterogeneous architecture involving many-core accelerators. On that supercomputer, the spectral element dynamical core of CAM has been refactored to finish 30-min prediction workload in 14.3s for a 3-km mesh simulation [17]. With graphics processing units (GPUs), some studies developed GPU-accelerated microphysics scheme of WRF and radiation parameterization of CAM, and obtained speedup of 70× and 14× [30, 31], respectively. The COnsortium for Small-scale MOdeling (COSMO) regional atmospheric model has been used to perform a near-global simulation with 930-m mesh using 4,888 GPUs, achieving 0.65 SDPH [18]. Some studies implemented the chemical kinetics scheme of the regional model on GPUs and obtained a speedup of 8.5× [32].

In this study, a non-hydrostatic global atmospheric modeling system with online integrated aerosol feedbacks (iAMAS) is established and optimized on the new Sunway supercomputer platform of China to significantly increase the computational efficiency and reduce the I/O cost. With the optimized iAMAS, the global 3-km resolution atmospheric simulation with online integrated natural aerosol feedbacks was scaled to 39,000,000 processor cores on the new Sunway supercomputer of China with heterogeneous architecture, and achieved the speed of 0.82 simulation day per hour (SDPH) with routine I/O that is the fastest computational speed ever reported for the similar type of experiment. This achievement finally enables us to perform the 5-day global weather forecast at 3-km horizontal resolution with online integrated natural aerosol impacts.

## 2. Methods

Below, we present the detailed processes of establishing iAMAS on the new Sunway Supercomputer, the optimization of massively parallelization of the model, and conducting global uniform 3-km numerical experiments to demonstrate the results of meteorology-aerosol online integrated weather forecast and natural aerosol impacts.

**2.1 iAMAS on the new Sunway Supercomputer**

The iAMAS model is established based on the MPAS-Atmosphere dynamic core that was developed for a fully compressible non-hydrostatic atmosphere, which discretizes the computational domain horizontally on a C-grid staggered unstructured Voronoi mesh using finite-volume formation [4]. The fully compressible non-hydrostatic equations are cast in terms of geometric-height hybrid terrain-following coordinate, and the solver applies the split-explicit time integration scheme. The time-integration scheme employs a 3rd-order Runge-Kutta (RK3) method and explicit time-splitting technique [33]. Recently, the MPAS-Atmosphere dynamic core has been widely used in different atmospheric models to address many important scientific questions [16, 19, 34-37].

For physics suite, iAMAS primarily treats sub-grid phenomena such as clouds, precipitation processes, longwave and shortwave radiation, and turbulent mixing. The physics parameterizations of iAMAS are taken from the advanced research Weather Research and Forecasting (WRF) model [2] and re-structured to be compatible with the architecture of the new Sunway Supercomputer.

For aerosol related suite, iAMAS includes the processes of online emission, advection, diffusion, vertical turbulent mixing, dry deposition, gravitational settling, and wet scavenging. The 10-bin size distribution parameterization of aerosol particles is adopted ranging from ~0.04 um to 40 um. In the experiments of this study, there are total of 123 natural aerosol related tracers (dust and sea salt) involved in these processes compared to 8 meteorology related scalars. Aerosols are transported with the high-order transport operator every time step that satisfies the consistency requirement on irregular Voronoi (hexagonal) meshes [38]. Aerosol optical properties are computed [39, 40] and coupled with the Rapid Radiative Transfer Model (RRTMG) [41, 42] for both shortwave and longwave radiation. Droplet activation and ice nucleation of aerosols are calculated following [43] and [44, 45], respectively, and then coupled with the Thompson cloud microphysics scheme [46]. More details about the aerosol related processes in iAMAS will be published in a separate study.

In general, the inclusion of aerosol suite can significantly reduce the computational efficiency compared to the meteorology-only simulation (Figure S1 in the supporting material). To simulate appropriately the atmospheric aerosols and their interactions with meteorology at the resolution of a few kilometers with acceptable computational cost, the scalability of iAMAS can be a vital issue for achieving acceptable performance on large-scale computing systems. Therefore, in this work, iAMAS was established on the new Sunway Supercomputer to be compatible with its heterogeneous many-core architecture. It includes a large codebase for more than 130 thousand lines of code (LOCs). More specifically, iAMAS is characterized by three phases: the physics (90 thousand LOCs), and the aerosol (31 thousand LOCs), the dynamics (16 thousand LOCs). These three phases are executed in turn during each simulation time-step.

**2.2 Massive parallelization of iAMAS on the new Sunway Supercomputer**

In order to solve the I/O bottleneck of the traditional atmospheric model and to significantly optimize the computational efficiency of the main kernels in the model so that the iAMAS simulation can be conducted on extreme-scale supercomputer such as the new Sunway Supercomputer, this study modifies the parallel I/O framework and uses a multi-level parallel optimization strategy corresponding to the many-core heterogeneous architecture of the new Sunway Supercomputer.

The new Sunway supercomputer is the new generation of Chinese home-grown supercomputer that inherits and develops the architecture of Sunway TaihuLight [47], based on new Sunway high-performance heterogeneous many-core processors SW26010 pro and interconnected network chips. The processor (Figure S2 in the supporting material) includes six core groups (CGs). Each CG integrates one management processing element (MPE) and one computing processing element (CPE) cluster of $8 \times 8$ CPEs that are connected through a ring network. Also, the MPE adopts self-developed SW64 instruction set with instruction cache and two-level caches. Each CPE has its own instruction cache and data storage that can be configured as a fully user-controlled local data memory (LDM) or can be configured partly as a hardware automatically managed data cache. Such a master-slave heterogeneous architecture different from the widely used X86 CPU (Central Processing Unit) requires designing the iAMAS system to take the advantage of its unique hardware.

2.2.1 New parallel I/O method

Global convection-permitting horizontal resolution (3-km mesh size) atmospheric simulation with iAMAS requires reading large input data (3 TB per time in the case of this study, see Section 2.3) during the initialization phase, and writing a large number of distributed arrays (1.1 TB per time in the case of this study) to disk every 3 simulation hours for the analysis afterwards during the real-world forecast. Besides, the checkpoint data (5.7 TB per time) also need to be written and read to restart the experiment in case of the failure of this large-scale supercomputer system. Also, a supercomputer's I/O subsystems are typically slow as compared to its other parts. Furthermore, high resolution simulation often requires many processes to improve computational speed, but I/O with many processes at the same time would increase the overall simulation time significantly. Parallel I/O (PIO) Library [48] is often used to provide a possible solution to release the stress of intensive I/O on file system. With PIO, a small number of processes are often selected manually responsible for all I/O, while the other processes obtain the required data from the I/O process through Message Passing Interface (MPI) communication instead of reading or writing directly from or to the disk. It is in $N - M - 1$ mode that M processes are selected as I/O processes among all N processes, and these M processes read/write one file at the same time as shown in the left panel of Figure 2a. Note that M cannot be set small because of the memory limitations when the amount of data becomes huge such as in global high-resolution simulation. However, large M would result in too many processes accessing the same file at the same time and thus leading to I/O contention problems.

Therefore, this study designs a new I/O framework called Fragmented PIO (F-PIO) that encapsulates original PIO, implementing simultaneous reading or writing of multiple files, and turning I/O pattern into an $N - M \times S - M$ pattern. In this pattern, M communicators are generated instead of original one communicator in PIO, and $M \times S$ processes are selected manually as I/O processes to read or write M files at the same time, where S processes read or write only one file and do not interact with other groups of I/O processes. The original single large input or output file is divided into M small files according to the cells assigned to the corresponding processes with pre-prepared static load balancing distribution file. Each file stores the data needed by N/M processes and these data are stored sequentially in the file according to the process number to ensure the data locality. In addition, the data needed by each process is stored in only one file, and there is no need to access other files during I/O. The right panel in Figure 2a illustrates the I/O pattern of F-PIO embedded in iAMAS. With the F-PIO method each input or output file only stores the variable values on the cells, edges, or vertices of the unstructured grid of iAMAS that are assigned to the corresponding process and its communication domain, which results in the order of cell indices within each file does not match the order of global cell indices. Therefore, an index mapping file is needed to remap the order of local cell of each input and output file to the order of global cell for the analysis afterwards. The F-PIO method developed in this work greatly reduces the I/O time of iAMAS when conducting real-atmosphere forecasts with high resolution, and thus solves the I/O bottleneck and improves significantly the scalability of iAMAS.

2.2.2 Multi-level optimization

As discussed above, one major challenge of including aerosol suite is the significant number of chemical species added that reduces the computational efficiency and increases the I/O burden. Therefore, the model is structured to allow multi-dimension (e.g., through horizontal, vertical, and field dimensions) parallelism (Fig. 2b). With this method, the multi-level parallelism offered by the heterogeneous architecture can be fully utilized.

First, for the dynamical kernels of iAMAS (i.e., MPAS-Atmosphere dynamical core), there are mainly three types of loops for mesh indices (cell, edge, vertex). Therefore, the loop bodies with the same loop steps are merged, which can increase the cache hit rate and reduce the overhead of swapping data in and out of the cache taking advantage of data locality. In addition, loop fusion can also lower the time of memory access by reducing data transmission between MPEs and CPEs. In addition, the loop bodies of dynamical transport procedures are embedded with neighboring communications when processes deliver a small amount of data to other processes, which occurs frequently and causes a lot of overhead. To reduce the overhead of dynamical processes, the multiple communications from all loop steps are aggregated into a single communication to make full use of the communication bandwidth by first decomposing a loop step into three parts (Fig. 2b): pre-communication, communication, and post-communication, and then saving the results of pre-communication computations needed by post-communication. More importantly, this code design aggregates the pre- and post-communication computations to ensure that enough computational tasks can be loaded onto the CPEs to make them computationally intensive and significantly increase the speedup ratio of these kernels after applied to the large-scale system.

Second, this study performs a compilation-guidance optimization to the code with a simple computational pattern in the dynamical core modules using the SWACC compiler based on the OpenACC 2.0 standard. This optimization method enables the distribution of the computational tasks of the mesh cells to CPEs to achieve thread-level parallelism. The optimization using OpenACC does not change the structure of the code, but only adds guidance statements to assign computational tasks to CPEs for computation.

The computational patterns of procedures in the physics and aerosol suites of iAMAS are more complicated and have deep function call stacks, which is not suitable for SWACC compilation guidance optimization. Therefore, this study designs these modules for the computational tasks of cells assigned to each process using the Athread interfaces on this new Sunway Supercomputer platform and partitions those tasks to CPEs to achieve thread-level parallelism.

Third, the computation in each CPE mainly focuses on the vertical model layer within each cell, and the computation and memory access patterns of each vertical layer within the cell are the same. Therefore, this study performs manual vectorization supported by CPE's 512-bit Single Instruction Multiple Data (SIMD) vector instruction for acceleration. The SIMD vector instruction not only reduces power consumption but also improves data-level parallelism (Fig. 2b).

Lastly, the dimensions of fields to be processed in the physics and aerosol modules are generally large for global high-resolution simulation. Therefore, the massive memory is required and poses a great challenge for the CPEs with only 256KB for local storage.

Compared with the previous generations of CG storage architecture, the local data storage of the CG of this new Sunway Supercomputer becomes configurable, which allows part of the data storage to be configured as a hardware automatically-managed local data cache. Therefore, this study sets 32KB of local storage as cache and others as LDM to satisfy the huge demand for local variables. Moreover, a combination of direct memory access (DMA) and the cache is used to reduce the latency when CPEs access the memory.

2.3 Numerical simulation

With the established iAMAS model on the new Sunway Supercomputer, the period with an atmospheric river (AR) event landing on the West United States (U.S.) (February 5-8 of 2015) is selected for conducting simulation to demonstrate the importance of including aerosol impacts on weather forecasts. This is, to the best of our knowledge, the first work to forecast the AR event with natural aerosol impacts using a global non-hydrostatic atmospheric model at a horizontal resolution of several kilometers. Traditionally, AR forecasting used global atmospheric model at relatively coarse-resolution for large-scale motions and applied a regional downscaling model to obtain high-resolution feature of meteorological fields such as precipitation [49], which has to face the challenges such as the inconsistency in physical processes between global and regional models that may introduce additional errors.

In this work, by using iAMAS on the new Sunway Supercomputer, the global weather forecast at 3-km with the online integrated natural aerosol impacts is achieved with good time-to-solution. The forecast is initialized with the ERA5 reanalysis data ($0.25° \times 0.25°$), which have 37 vertical levels ranging from 1000 hPa to 1 hPa. During the model integration, sea surface temperature and sea ice cover are updated every six hours. The two-month simulation with the same configuration at the uniform 120 km horizontal resolution is conducted to provide chemical initial condition (natural dust and sea salt) for 3-day warm-up experiments that start at UTC 00:00 on 1 February, 2015. To avoid the impacts of initial condition, three ensemble forecasts (three hours difference in starting forecast time) are carried out from UTC 00:00, 4 February 2015 with the chemical initial condition from the 3-day warm-up experiments, and the average of ensemble forecast results for Feb. 5-8 is analyzed and shown. Although some studies suggested that a longer spin-up time (more than one day) might be needed for global simulations at horizontal resolutions of a few kilometers [50], one-day spin-up time may be acceptable in this study and was also used in other studies of global high-resolution simulations [19, 51]. In particular, the modeling results after the spin-up time in this study are quite reasonable compared to the stational observations and the reanalysis (see Section 3.2). The impact of spin-up time on the simulated results deserves further investigation but is beyond the scope of this study.

To demonstrate and evaluate the performance of global 3-km simulation, this study conducts three sets of forecasts with different mesh configurations, i.e., global uniform 60km mesh (U60KM, 163842 mesh cells), variable 4-60km mesh with the refinement of 4km mesh centering at the western coast of U.S. (V4KM, 785410 mesh cells), and global uniform 3km mesh (U3KM, 65536002 mesh cells). Besides the results from global 3km forecast, the difference between U60KM and U3KM is also shown. The number of vertical layers is 56 for all experiments, and the layer thickness gradually increases from 60 m near the surface to 1000 m at 30 km (the model top). The time step for processes of dynamics, physics, and aerosol is 15 s, 20 s, and 300 s for U3KM, V4KM, and U60KM experiments, respectively. All experiments use the Grell-Freitas cumulus scheme which is the scale-aware convective parameterization available in the current version of iAMAS, the Thompson microphysics scheme, the YSU planetary boundary layer scheme, the Noah land surface scheme, and the RRTMG short and longwave radiation schemes. The parameterizations are not tuned for each experiment. Besides these standard experiments, another set of experiments without natural aerosol emissions are also conducted to show the impact of natural aerosols on global weather forecast.

## 3. Results and discussion

In this section, we demonstrate the computational efficiency and parallel scalability of iAMAS on the new Sunway Supercomputer. In addition, the impacts of global convection-permitting horizontal resolution and aerosols on weather forecast are also demonstrated.

### 3.1 Performance results

To evaluate the performance of the global atmospheric simulation with the established iAMAS on the new Sunway Supercomputer, the necessity of I/O improvement for the computation with the large-scale system is first proved. The need of optimizing the computational hotspot part of the iAMAS model in order to take the advantage of the many-core architecture is demonstrated. The performance aspects of iAMAS simulations for different problem sizes of U60KM, V4KM, and U3KM are measured. Lastly, the scalability of U3KM with the largest problem size up to 600,000 CGs is reported.

3.1.1 I/O optimization for iAMAS

During the initialization of iAMAS, the model needs to assign global mesh cells to each MPI task. The set of cells assigned to an MPI task is referred to as a block. The model then divides the global arrays into many distributed but adjacent arrays that can be assigned to the corresponding processes, which is referred to as the bootstrapping step. When reaching the scale of tens of thousands of processes, the bootstrapping step would consume unacceptable time. To demonstrate the improvement of I/O efficiency, the time consumed by a single variable (water vapor mixing ratio) for the U3KM experiment is examined with the process number ranging from 30,000 to 120,000. We only test the system scale up to 120,000 in order to finish the test in an acceptable time. When using a single file, the decomposed time of one field with PIO increases with the enlarging process number from 76.6s of 30,000 processes to 576.6s of 120,000 processes. On the contrary, the initialization time with F-PIO decreases with the enlarging process number from 0.11s of 30,000 processes to 0.027s of 120,000 processes. This is because, as discussed above, F-PIO not only reduces the number of processes involved in communication but also decomposes the variables more efficiently and diminishes the size of single file, all of which contribute to lowering the overhead of communications and initialization time. The results achieved for the acceleration of I/O initialization with PIO and F-PIO are summarized in Table S1 in the supporting material. Please note that the time discussed above is only for reading one field out of hundreds of initial fields for U3KM.

Figure 3a shows the overall time of reading and writing files for the problem sizes of U60KM, V4KM, and U3KM using PIO and F-PIO. The U60KM simulation runs with 240 processes, the V4KM simulation runs on 12,000 processes, and the U3KM simulation runs with 30000 processes. With PIO U60KM takes 10.1s and 13.2s to finish the reading of the input file and one single output procedure, respectively, while with F-PIO it takes only 1.1s and 5.1s correspondingly. For the V4KM problem size, with 95.2s for PIO and 1.7s for F-PIO for the input loading, and with 46.0s for PIO and 3.5s for F-PIO for the single output procedure. The acceleration is more evident for the U3KM problem size. PIO takes 709.8s to finish loading input file, which is only 21.2s with F-PIO. To output 1.2 TB of data, F-PIO needs only 27.3s rather than 7723.5s with PIO. This proves that the iAMAS simulation with F-PIO is significantly faster than with PIO.

3.1.2 Kernel optimization for iAMAS

Figure 3b shows the relative computational percentage of the nine groups of most time-consuming kernels of iAMAS, which occupy most (84%) of the entire integration time during the simulation when using MPE only. Three of these groups of kernels are aerosol-related processes (i.e., optical, drydep and wetscav) that consume ~24% of the entire integration time. Two kernels related to tracer transport (i.e., transport_1 and transport_2) consumes ~47% of the entire integration time, mainly due to the exponential increase in the amount of chemical tracers. As discussed in Section 2.2.2, some efforts are made to optimize the code structure to take the advantage of new Sunway heterogeneous architecture. Figure 3c shows the speedup of these kernels for using SW26010 pro processors. The performance speedups are sitting on top of each bar, with the MPE performance being the baseline. For each kernel, the results demonstrate the evident accelerations with the 64 CPEs. Compared with the performance using one MPE process, scaling the performance of 64 CPEs would improve the performance by 3× to 13×. Particularly for the aerosol-related kernels, the speedups are significant. The speedups of transport_1 and transport_2 are 6.2× and 7.1×, which reduces significantly the computational percentage of these two kernels from 47% to 17%. These results demonstrate that designing the kernels specifically to the heterogeneous architecture, particularly for the tracer transport-related kernels, can significantly improve the computational efficiency of global meteorology-aerosol online coupled simulation on the new Sunway Supercomputer.

3.1.3 Scaling result for iAMAS

As shown in Figure 4a, with the number of processes increasing from 120 to 3840 for the U60KM simulation, we can observe reasonable performance benefits from the acceleration with OpenACC and Athread by comparing MPE and CPE performance. The fastest simulation speed achieved for U60KM is 26.3 SDPH when using 3840 processes in our test. Figure 4b indicates the similar acceleration for the V4KM simulation, and the simulation speed of V4KM can reach 2.67 SDPH with 32000 processes. Figure 4c shows the scaling result of U3KM. It indicates that the performance improves from 0.054 SDPH to 0.82 SDPH when enlarging the system scale from 30,000 processes to 600,000 processes. Compared to the simulation speed range (0.10-0.65 SDPH; [16-18, 20, 22, 24]) from previous studies of the global non-hydrostatic atmospheric simulations at kilometer-level mesh, this study also represents a significant improvement in computational performance achievement. In particular, this simulation speed measured in this study is for meteorology-aerosol online coupled forecast and includes routine I/O (46GB every simulation hour, 1.1TB every 3 simulation hours, 5.7TB every 3 simulation hours).

In addition, this study also presents a significant scaling improvement of MPAS dynamical core, compared with one previous study that also used MPAS as the dynamic core and conducted a global 3-km simulation with 24576 CPU nodes (393,216 processes, 167 cells per task) [16]. They reported the simulation speed of 0.32 SDPH. The experimental results in this study show that the iAMAS established on the new Sunway Supercomputer achieves strong scalability for different scales of processes for U3KM, which was obtained by measuring the time of a single integration step. The number of new Sunway processes (or CGs) increases from 30,000 to 600,000 CGs with the parallel efficiency still maintaining ~76%. Please note, although the overall communication time during the simulation increases with the system scale, the increasing of system scale decreases the number of cells assigned to each process and thus its integration time and communication effort to other processes (the simple relation between the number of first halo cells $N_h$ and the number of owned cells $N_o$ assigned to each process can be described as $N_h \sim 4\sqrt{\pi N_o}$ [16]), which makes the parallel efficiency does not drop too much.

**3.2 Weather Forecast with natural aerosol impacts**

AR is a prominent feature of the global water cycle and a typical weather event. On long-term average, 20%–50% of annual precipitation over the western US is attributed to six to seven AR events in the cold season that produce heavy precipitation [52]. Flooding is more likely to occur [53] when an AR makes landfall on preexisting snowpack and high antecedent soil moisture conditions. Therefore, the period of an AR event landing on the West U.S. on February 4-8 of 2015 is selected for weather forecast. The results averaged for this period are analyzed to demonstrate the impacts of global convection-permitting resolution and natural aerosol on weather forecast. Please note again, the main purpose of this study is to demonstrate an establishment of a meteorology-aerosol online coupled non-hydrostatic global atmospheric modeling system on the new Sunway supercomputer platform, which allows us to perform the global weather forecast at a few kilometers horizontal resolution with online natural aerosol impacts. The detailed investigation of the mechanisms driving the impacts of resolution and aerosol on the weather fields is beyond the scope of this study and will be presented in future studies.

Figure 5 shows the global distributions of rainfall, temperature at 850 hPa, and wind speed at 850 hPa averaged during the period (Feb. 5-8) from the U3KM experiment. U3KM captures the spatial structure of the AR event and associated precipitation over the western U.S. The result also shows relatively high precipitation over the tropics. In general, precipitation is higher over the ocean than the land. The difference between U3KM and U60KM (Fig. S3 in the supporting material) shows that U3KM produces significant different precipitation from U60KM over the tropics and also the mid-latitude such as the western U.S., North India, and Northwest Pacific. Fig. 5b shows the root mean square error (RMSE) of simulated rainfall from the U3KM and U60KM experiments against the hourly precipitation dataset from the National Climate Data Center (NCDC) of the United States [54]. The RMSE is calculated against the data from 226 observational sites over the western US as shown in Fig. 6 to mainly reflect the AR associated rainfall. It is evident that U3KM improves the prediction of AR associated rainfall compared with U60KM, reflected by smaller RMSE on Feb. 7-8$^{th}$. For temperature at 850 hPa (Fig. 5c), U3KM simulates relatively high temperature between 30°N-30°S, particularly over the land. In general, U3KM produces higher temperature than U60KM over East Asia except lower temperature over Northwest China. Fig. 5d shows the RMSE of simulated temperature at 850 hPa from the U3KM and U60KM experiments against the final operational global analysis dataset from NCEP (NCEP-FNL) with 6-hr interval and global coverage at 1°×1° horizontal resolution [55]. It is evident that U3KM reduces the prediction biases of temperature at 850 hPa compared with U60KM during the simulation period. In addition, U3KM also significantly reduces the prediction biases of wind speed at 850 hPa than U60KM (Fig. 5e-f) against the NCEP-FNL dataset globally. The evaluations against the hourly reanalysis dataset of ERA5 for temperature and wind show similar results (not shown).

Global distributions of natural aerosol impacts on precipitation show largest impacts on precipitation over the tropics, corresponding to the heavy precipitation over that region (Fig. S4 in the supporting material). In the mid-latitude, the impacts on the heavy precipitation are mainly associated with the AR. Figure 6 shows the spatial distributions of the difference of rainfall at the observational stations from the NCDC over the western United States between the U3KM experiments with and without aerosol impacts and between the U3KM experiment without aerosol impacts and the observations. For the AR associated precipitation at the western coast of the U.S., natural aerosols tend to enhance precipitation over the South and reduce it over the North (Fig. 6a). Figure 6b shows the forecast biases in precipitation from the U3KM experiment without aerosol impacts. It indicates that without aerosol impacts U3KM tends to underestimate (overestimate) the AR associated precipitation over the southwestern (northwestern) the U.S. Therefore, it illustrates that aerosol impacts improve the AR associated precipitation during the simulation period. This aerosol-induced shift of AR rainband may be resulted from the change of large-scale circulation primarily due to aerosol-radiation interaction. As mentioned above, the detailed investigation of the mechanisms of aerosol impacts will be investigated in future studies.

## 4. Conclusion and implications

In this study, a meteorology-aerosol online coupled atmospheric modeling system (iAMAS) is established for a supercomputer system with heterogeneous architecture. Through using aggressive and finer-grained optimization, applying manual vectorization, and innovating the Fragmented PIO method with this unique architecture, the computational efficiency of iAMAS is significantly increased and the I/O cost of iAMAS is reduced. The computational speeds of key time-consuming kernels of iAMAS are increased by 3× to 13×. The iAMAS modeling system is scaled to the 600,000 CGs (39,000,000 processor cores) and achieves the speed of 0.82 SDPH for global 3-km non-hydrostatic weather forecast, which is the fastest speed first-ever reported, particularly considering the online coupled meteorology-aerosol processes and with routine I/O. This enables us to perform a 5-day global weather forecast at 3-km resolution with the impacts from natural aerosols. The analysis with observations and reanalysis datasets demonstrates the benefits of increasing the spatial resolution to a few kilometers with online coupled aerosol impacts, which significantly improve the global weather forecast of precipitation and temperature during the simulation period in this study. Nevertheless, the experiments for more weather events and longer period are needed to illustrate thoroughly the stability and advantage of this modeling system in terms of convection-permitting resolution and aerosol impact. The mechanisms of aerosol impacts in such a system also deserve further investigation.

Our efforts and results support the promising future of real-world non-hydrostatic global weather and air quality fully coupled forecast system with the high horizontal resolution of a few kilometers. This is particularly important for the regions where natural and anthropogenic aerosols may have significant impacts on weather systems. For example, over Asia, both natural and anthropogenic aerosols are abundant and previous studies have shown their impacts on Asian weather and climate. Investigation of numerical forecast at high horizontal resolution of a few kilometers for typical severe weather events such as heat wave and Mei-yu and Typhoon rainfall with online aerosol impacts is of particular interest in future. This modeling system can also be useful for better estimating anthropogenic forcing for climate change such as the forcing of greenhouse gases and aerosols. Some modeling sensitivities related to global convection-permitting resolution deserve further investigation, including the scale-awareness of parameterizations and impact of spin-up time.

**Conflict of interest**

The authors declare that they have no conflict of interest.

**Acknowledgements**

This research was supported by the Strategic Priority Research Program of Chinese Academy of Sciences (Grant XDB41000000) and the USTC Research Funds of the Double First-Class Initiative (Grant YD2080002007), the National Natural Science Foundation of China (Grant 91837310, 42061134009, 41775146). The study used the computing resources from the Qingdao Supercomputing and Big Data Center, the High-Performance Computing Center of University of Science and Technology of China (USTC), and the TH-2 of National Supercomputer Center in Guangzhou (NSCC-GZ).

**Author contributions**

Chun Zhao and Hong An conceived the study and led the overall scientific questions. Jun Gu, Tao Fang, Xiaoyu Hao, Jiawang Feng developed the model and conducted the experiments. Jun Gu, Tao Fang, Xiaoyu Hao, Chun Zhao, Hong An, Junshi Chen, Jiawang Feng performed the data analysis. All authors are involved in discussing and writing the paper.

**Reference**
[1] Charney JG, Eliassen A. A Numerical Method for Predicting the Perturbations of the Middle Latitude Westerlies. Tellus, 1949, 1: 38-54
[2] Skamarock WC, Klemp JB, Dudhia J, et al. A description of the Advanced Research WRF version 3. NCAR Technical note-475+ STR. 2008,
[3] Neale RB, Chen C-C, Gettelman A, et al. Description of the NCAR community atmosphere model (CAM 5.0). NCAR Tech Note NCAR/TN-486+ STR, 2010, 1: 1:12
[4] Skamarock WC, Klemp JB, Duda MG, et al. A Multiscale Nonhydrostatic Atmospheric Model Using Centroidal Voronoi Tesselations and C-Grid Staggering. Mon Weather Rev, 2012, 140: 3090-3105
[5] Satoh M, Matsuno T, Tomita H, et al. Nonhydrostatic icosahedral atmospheric model (NICAM) for global cloud resolving simulations. Journal of Computational Physics, 2008, 227: 3486-3514
[6] Kühnlein C, Deconinck W, Klein R, et al. FVM 1.0: a nonhydrostatic finite-volume dynamical core for the IFS. Geosci Model Dev, 2019, 12: 651-676
[7] Randall DA. Beyond deadlock. Geophys Res Lett, 2013, 40: 5970-5976
[8] Huang X, Ding A. Aerosol as a critical factor causing forecast biases of air temperature in global numerical weather prediction models. Science Bulletin, 2021,
[9] Bender FAM. Aerosol Forcing: Still Uncertain, Still Relevant. AGU Advances, 2020, 1:
[10] Bellouin N, Quaas J, Gryspeerdt E, et al. Bounding Global Aerosol Radiative Forcing of Climate Change. Rev Geophys, 2020, 58: e2019RG000660
[11] Rosenfeld D, Zhu YN, Wang MH, et al. Aerosol-driven droplet concentrations dominate coverage and water of oceanic low-level clouds. Science, 2019, 363: 599-+
[12] Myhre G, Samset BH, Schulz M, et al. Radiative forcing of the direct aerosol effect from AeroCom Phase II simulations. Atmos Chem Phys, 2013, 13: 1853-1877
[13] Miura H, Satoh M, Nasuno T, et al. A Madden-Julian Oscillation event realistically simulated by a global cloud-resolving model. Science, 2007, 318: 1763-1765
[14] Miyamoto Y, Kajikawa Y, Yoshida R, et al. Deep moist atmospheric convection in a subkilometer global simulation. Geophys Res Lett, 2013, 40: 4922-4926
[15] Skamarock WC, Park SH, Klemp JB, et al. Atmospheric Kinetic Energy Spectra from Global High-Resolution Nonhydrostatic Simulations. J Atmos Sci, 2014, 71: 4369-4381
[16] Heinzeller D, Duda MG, Kunstmann H. Towards convection-resolving, global atmospheric simulations with the Model for Prediction Across Scales (MPAS) v3.1: an extreme scaling experiment. Geosci Model Dev, 2016, 9: 77-110
[17] Fu HH, Liao JF, Ding N, et al. Redesigning CAM-SE for Peta-Scale Climate Modeling Performance and Ultra-High Resolution on Sunway TaihuLight. Sc'17: Proceedings of the International Conference for High Performance Computing, Networking, Storage and Analysis, 2017,
[18] Fuhrer O, Chadha T, Hoefler T, et al. Near-global climate simulation at 1 km resolution: establishing a performance baseline on 4888 GPUs with COSMO 5.0. Geosci Model Dev, 2018, 11: 1665-1681
[19] Judt F. Insights into Atmospheric Predictability through Global Convection-Permitting Model Simulations. J Atmos Sci, 2018, 75: 1477-1497

[20] Yashiro H, Terasaki K, Kawai Y, et al. A 1024-member ensemble data assimilation with 3.5-km mesh global weather simulations. In: Proceedings of the SC20: International Conference for High Performance Computing, Networking, Storage and Analysis, 2020. IEEE Press:

[21] Michalakes J, Govett M, Benson R, et al. AVEC Report: NGGPS Level-1 Benchmarks and Software Evaluation. 2015,

[22] Yang C, Xue W, Fu HH, et al. 10M-Core Scalable Fully-Implicit Solver for Nonhydrostatic Atmospheric Dynamics. Sc '16: Proceedings of the International Conference for High Performance Computing, Networking, Storage and Analysis, 2016, 57-68

[23] Stevens B, Satoh M, Auger L, et al. DYAMOND: the DYnamics of the Atmospheric general circulation Modeled On Non-hydrostatic Domains. Prog Earth Planet Sc, 2019, 6:

[24] Wedi NP, Polichtchouk I, Dueben P, et al. A Baseline for Global Weather and Climate Simulations at 1 km Resolution. J Adv Model Earth Sy, 2020, 12:

[25] Yang Q, Leung LR, Lu J, et al. Exploring the effects of a nonhydrostatic dynamical core in high-resolution aquaplanet simulations. Journal of Geophysical Research: Atmospheres, 2017, 122: 3245-3265

[26] Suzuki K, Nakajima T, Satoh M, et al. Global cloud‐system‐resolving simulation of aerosol effect on warm clouds. Geophys Res Lett, 2008, 35:

[27] Sato Y, Miura H, Yashiro H, et al. Unrealistically pristine air in the Arctic produced by current global scale models. Scientific Reports, 2016, 6: 1-9

[28] Baklanov A, Schlünzen K, Suppan P, et al. Online coupled regional meteorology chemistry models in Europe: current status and prospects. Atmos Chem Phys, 2014, 14: 317-398

[29] Zhang Y. Online-coupled meteorology and chemistry models: history, current status, and outlook. Atmos Chem Phys, 2008, 8: 2895-2932

[30] Mielikainen J, Huang B, Wang J, et al. Compute unified device architecture (CUDA)-based parallelization of WRF Kessler cloud microphysics scheme. Comput Geosci-Uk, 2013, 52: 292-299

[31] Kelly R. GPU Computing for Atmospheric Modeling Experience with a Small Kernel and Implications for a Full Model. Comput Sci Eng, 2010, 12: 26-32

[32] Linford JC, Michalakes J, Vachharajani M, et al. Multi-core Acceleration of Chemical Kinetics for Simulation and Prediction. Proceedings of the Conference on High Performance Computing Networking, Storage and Analysis, 2009,

[33] Wicker LJ, Skamarock WC. Time-splitting methods for elastic models using forward time schemes. Mon Weather Rev, 2002, 130: 2088-2097

[34] Hagos S, Leung LR, Yang Q, et al. Resolution and Dynamical Core Dependence of Atmospheric River Frequency in Global Model Simulations. J Climate, 2015, 28: 2764-2776

[35] Sakaguchi K, Leung LR, Zhao C, et al. Exploring a Multiresolution Approach Using AMIP Simulations. J Climate, 2015, 28: 5549-5574

[36] Zhao C, Leung LR, Park SH, et al. Exploring the impacts of physics and resolution on aqua-planet simulations from a nonhydrostatic global variable-resolution modeling framework. J Adv Model Earth Sy, 2016, 8: 1751-1768

[37]     Zhao C, Xu MY, Wang Y, et al. Modeling extreme precipitation over East China with a global variable-resolution modeling framework (MPASv5.2): impacts of resolution and physics. Geosci Model Dev, 2019, 12: 2707-2726
[38]     Skamarock WC, Gassmann A. Conservative transport schemes for spherical geodesic grids: High-order flux operators for ODE-based time integration. Mon Weather Rev, 2011, 139: 2962-2975
[39]     Fast J, Gustafson Jr W, Easter R, et al. Evolution of ozone, particulates, and aerosol direct forcing in an urban area using a new fully-coupled meteorology, chemistry, and aerosol model. J Geophys Res, 2006, 111: D21305
[40]     Barnard JC, Fast JD, Paredes-Miranda G, et al. Evaluation of the WRF-Chem" Aerosol Chemical to Aerosol Optical Properties" Module using data from the MILAGRO campaign. Atmos Chem Phys, 2010, 10: 7325-7340
[41]     Mlawer EJ, Taubman SJ, Brown PD, et al. Radiative transfer for inhomogeneous atmospheres: RRTM, a validated correlated‐k model for the longwave. Journal of Geophysical Research: Atmospheres, 1997, 102: 16663-16682
[42]     Iacono MJ, Mlawer EJ, Clough SA, et al. Impact of an improved longwave radiation model, RRTM, on the energy budget and thermodynamic properties of the NCAR community climate model, CCM3. Journal of Geophysical Research: Atmospheres, 2000, 105: 14873-14890
[43]     Ghan S, Laulainen N, Easter R, et al. Evaluation of aerosol direct radiative forcing in MIRAGE. Journal of Geophysical Research: Atmospheres, 2001, 106: 5295-5316
[44]     DeMott PJ, Prenni AJ, McMeeking GR, et al. Integrating laboratory and field data to quantify the immersion freezing ice nucleation activity of mineral dust particles. Atmos Chem Phys, 2015, 15: 393-409
[45]     Liu X, Penner J. Ice nucleation parameterization for global models. Meteorologische Zeitschrift, 2005, 14: 499-514
[46]     Thompson G, Field PR, Rasmussen RM, et al. Explicit forecasts of winter precipitation using an improved bulk microphysics scheme. Part II: Implementation of a new snow parameterization. Mon Weather Rev, 2008, 136: 5095-5115
[47]     Fu H, Liao J, Yang J, et al. The Sunway TaihuLight supercomputer: system and applications. Science China Information Sciences, 2016, 59: 1-16
[48]     Hartnett E, Edwards J. THE PARALLELIO (PIO) C/FORTRAN LIBRARIES FOR SCALABLE HPC PERFORMANCE. In: Proceedings of the 101st American Meteorological Society Annual Meeting, 2021. AMS:
[49]     Leung LR, Qian Y. Atmospheric rivers induced heavy precipitation and flooding in the western US simulated by the WRF regional climate model. Geophys Res Lett, 2009, 36:
[50]     Kajikawa Y, Miyamoto Y, Yoshida R, et al. Resolution dependence of deep convections in a global simulation from over 10-kilometer to sub-kilometer grid spacing. Prog Earth Planet Sc, 2016, 3: 16
[51]     Skamarock WC, Duda MG, Ha S, et al. Limited-Area Atmospheric Modeling Using an Unstructured Mesh. Mon Weather Rev, 2018, 146: 3445-3460
[52]     Dettinger MD, Ralph FM, Das T, et al. Atmospheric Rivers, Floods and the Water Resources of California. Water-Sui, 2011, 3: 445-478

[53]     Ralph FM, Coleman T, Neiman PJ, et al. Observed Impacts of Duration and Seasonality of Atmospheric-River Landfalls on Soil Moisture and Runoff in Coastal Northern California. J Hydrometeorol, 2013, 14: 443-459

[54]     Wuertz D, Lawrimore J, Korzeniewski B. Cooperative Observer Program (COOP) Hourly Precipitation Data (HPD), Version 2.0 (2018). NOAA National Centers for Environmental Information. doi:10.25921/p7j8-2170 [last access: Sep. 22, 2021]

[55]     NCEP FNL Operational Model Global Tropospheric Analyses, continuing from July 1999. Research Data Archive at the National Center for Atmospheric Research, Computational and Information Systems Laboratory, 2000. 10.5065/D6M043C6 [last access: Dec. 7, 2021]

Figure captions:

**Figure 1.** a) Atmosphere includes multiple scale processes from a few meters to thousands of kilometers that eventually affect our weather, air quality, and climate. High horizontal resolution is required for modeling these multiscale atmospheric processes; (b) Atmospheric model integrates a variety of fluid-dynamical, physical, chemical and geobiological procedures across different temporal and spatial scales. Atmospheric chemistry processes are sophisticated and important components of atmospheric model.

**Figure 2.** a) iAMAS with PIO (L) and F-PIO (R). (L) is in $N - M - 1$ mode that M processes access one single file at the same time, and (R) is in $N - M \times S - M$ mode that $M \times S$ I/O processes read or write M files with S processes access one file. b) Sketch-map of multi-dimension parallelism structure. (U) Communication decomposition (L) Multi-dimension parallelism.

**Figure 3.** a) Each I/O step time of different problem sizes. b) Computational percentage of the most time-consuming kernels. (c) Speedup with CPEs against MPE.

**Figure 4.** a) SDPH for the U60KM experiments with different numbers of processes. b) SDPH for the V4KM experiments with different numbers of processes. c) SDPH and Parallel efficiency for the U3KM experiments with different numbers of processes.

**Figure 5.** The global distributions of (a) rainfall, (c) temperature at 850 hPa, and (e) wind speed at 850 hPa averaged during the period (Feb. 5-8) from the U3KM experiment. (b), (d), (f) show the corresponding root mean square error (RMSE) of simulated rainfall, temperature, and wind speed from the U3KM and U60KM experiments against the NCDC observation and NCEP FNL reanalysis. The RMSE of rainfall is from the comparison with the hourly stational observations over the Western United States available from the NCDC as shown in Fig. 6a, while the RMSE of temperature and wind speed is from the comparison with the NCEP FNL reanalysis globally.

**Figure 6.** The spatial distributions of the difference of rainfall at the observational stations from the NCDC over the western United States (a) between the U3KM experiments with and without aerosol impacts and (b) between the U3KM experiment without aerosol impacts and the observations.

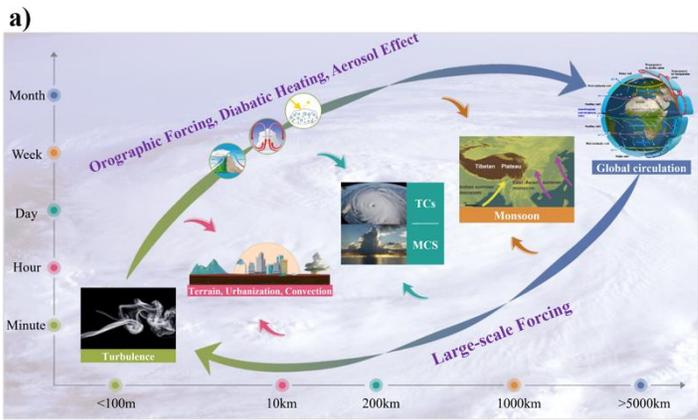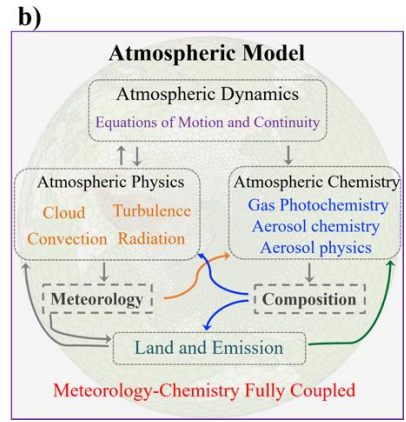

**Figure 1**

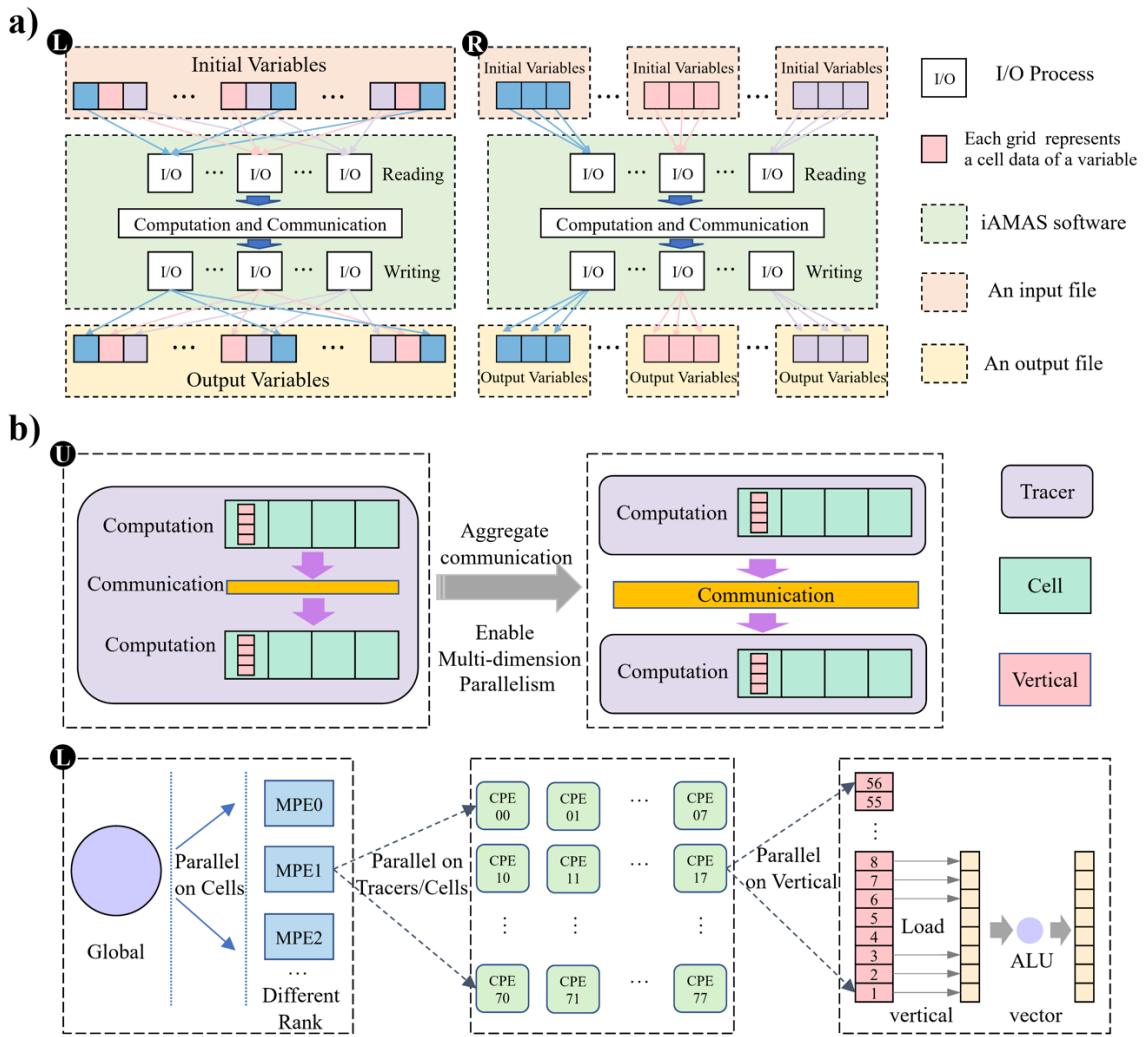

**Figure 2**

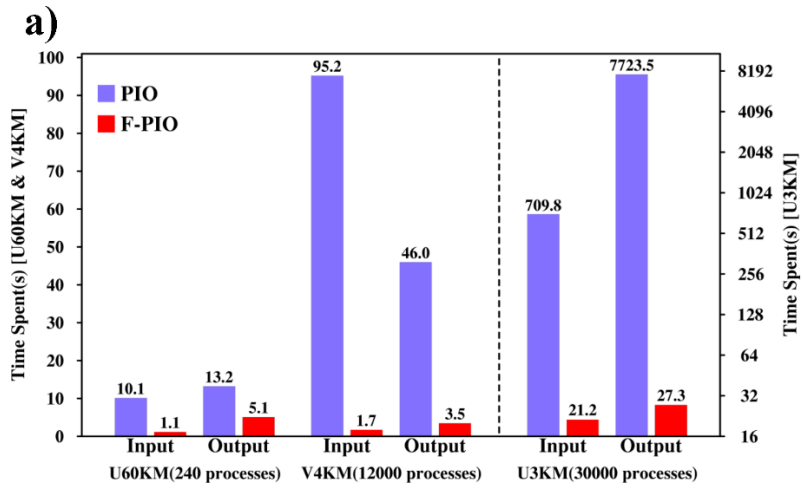

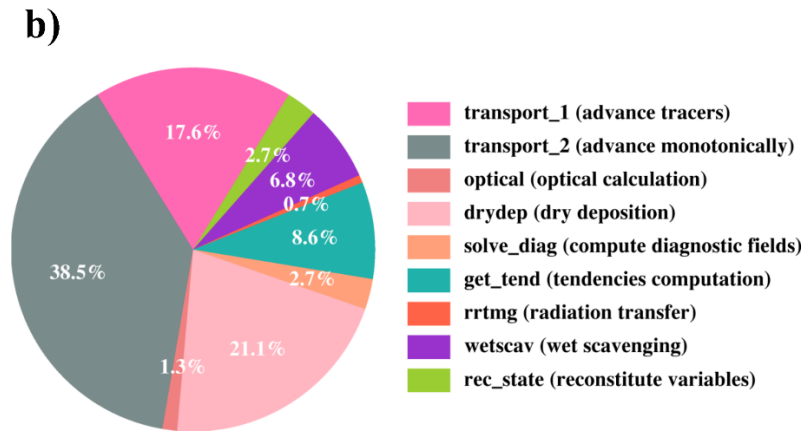

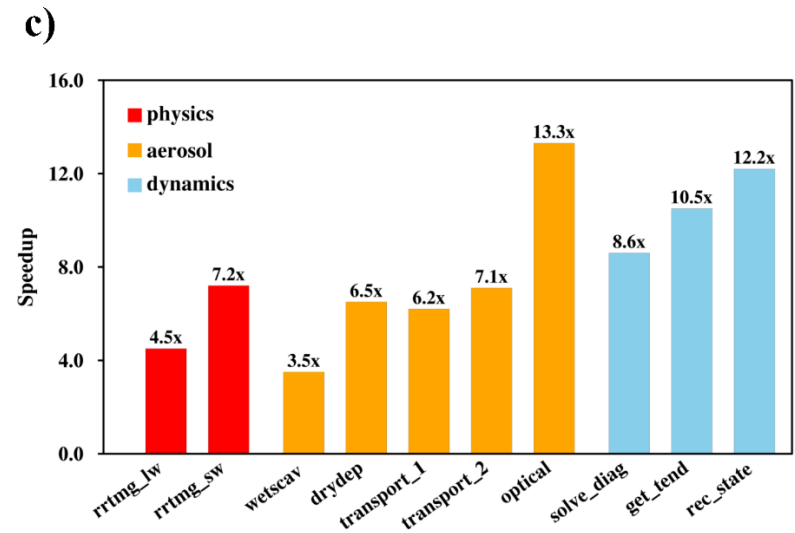

**Figure 3**

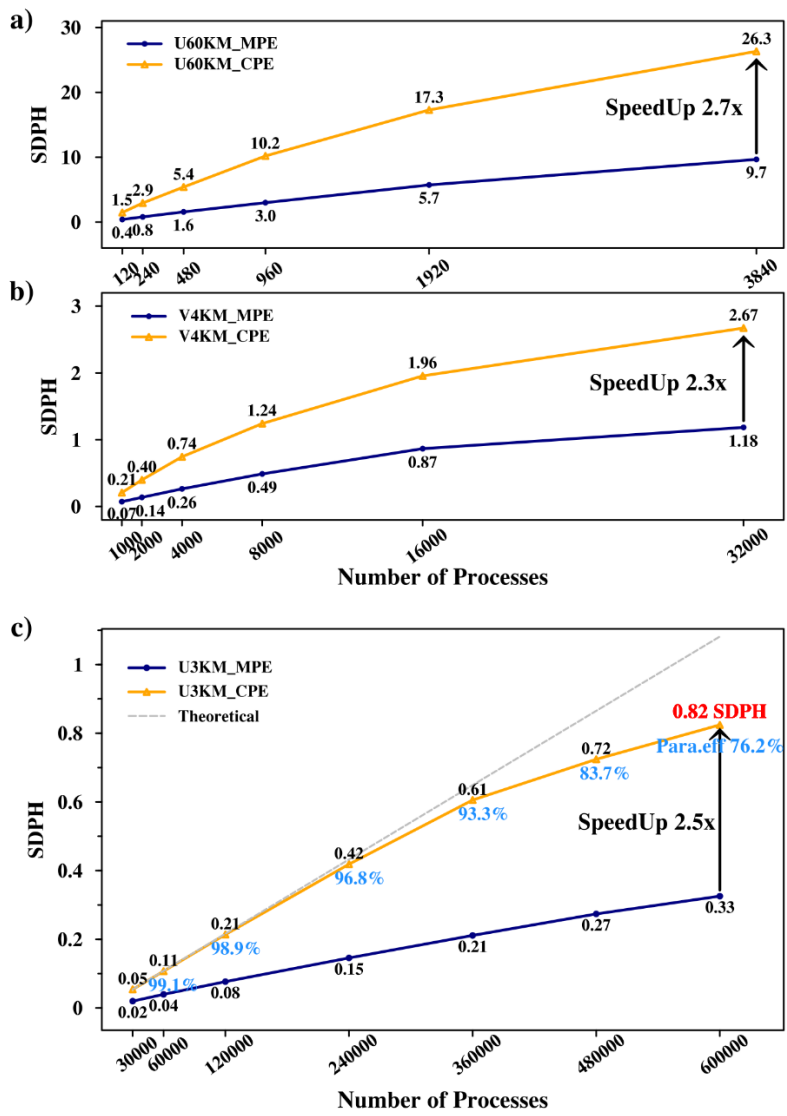

**Figure 4**

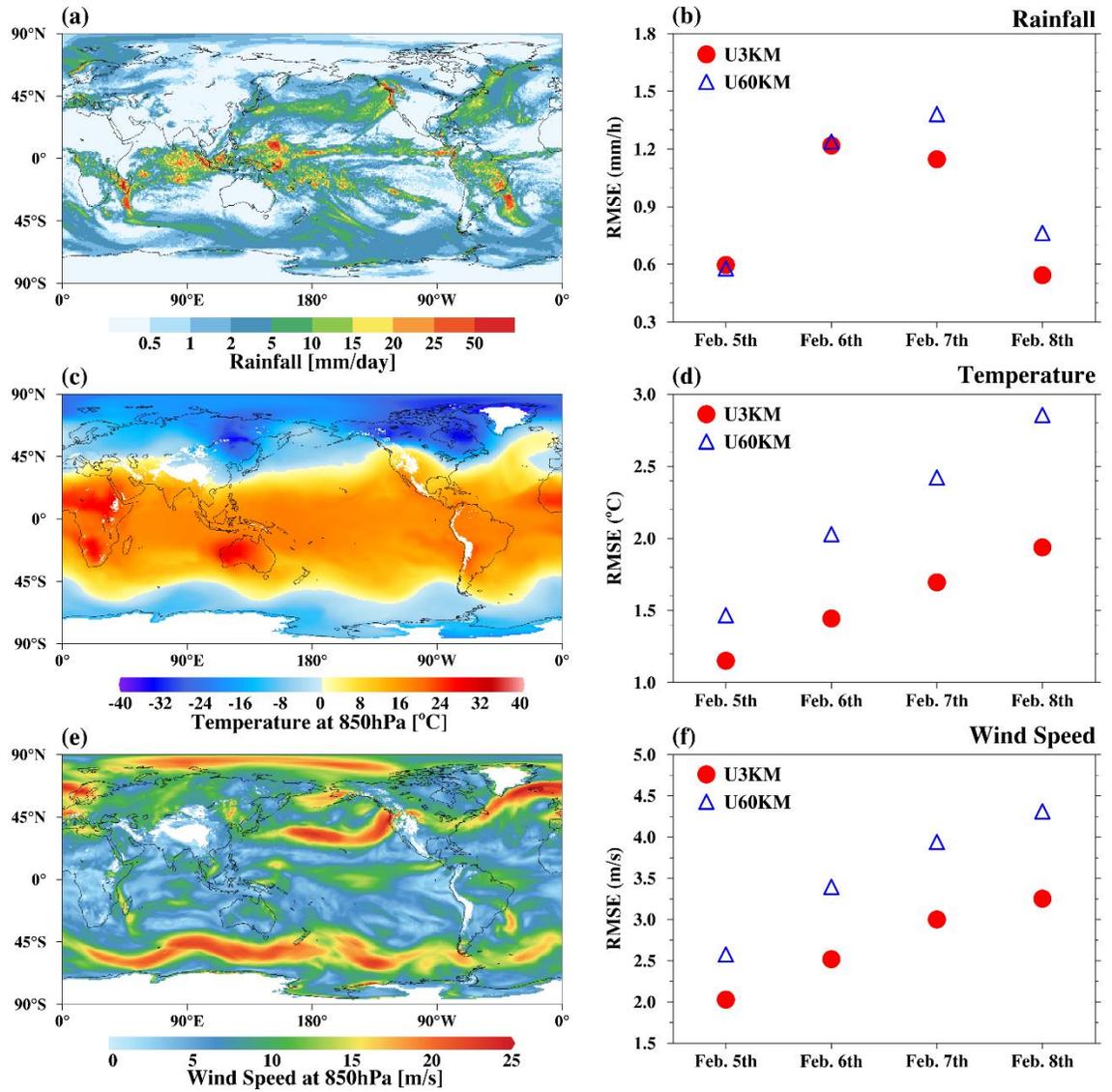

**Figure 5**

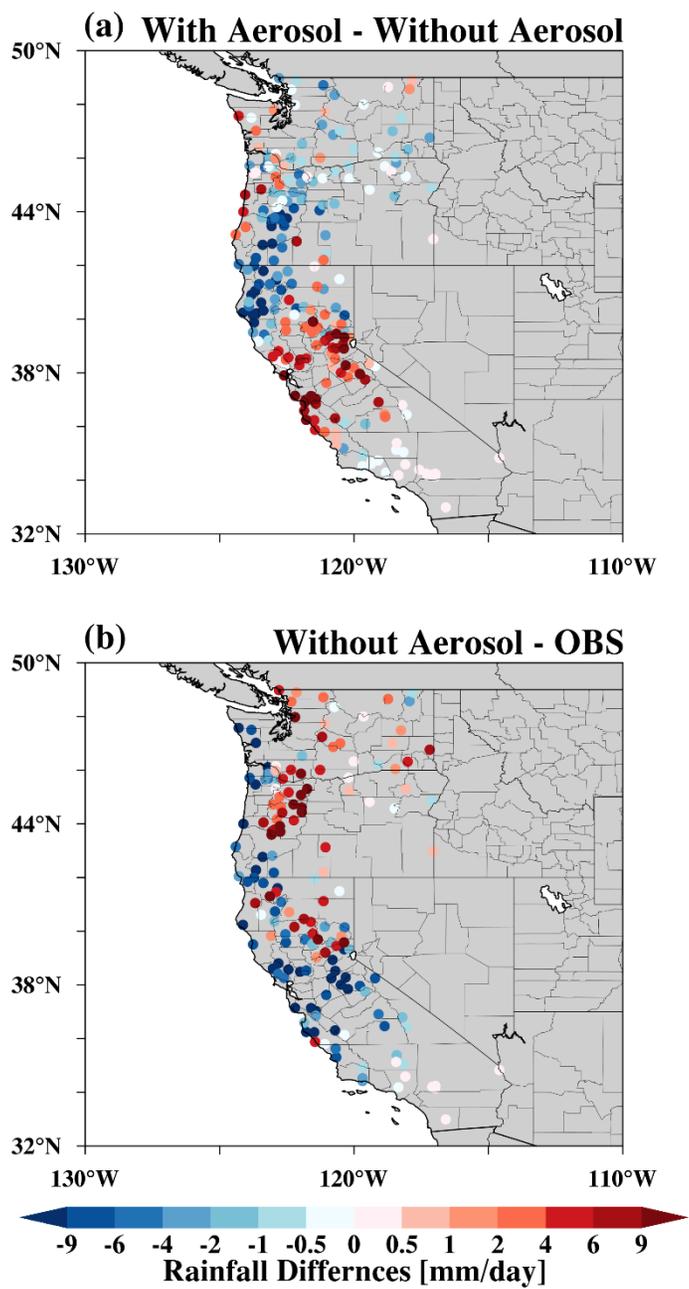

**Figure 6**

Supporting Information for

**Establishing a non-hydrostatic global atmospheric modeling system (iAMAS) at 3-km horizontal resolution with online integrated aerosol feedbacks on the Sunway supercomputer of China**

Jun Gu[1#], Jiawang Feng[1#], Xiaoyu Hao[2#], Tao Fang[2#], Chun Zhao[1,3,4 *], Hong An[2*], Junshi Chen[2], Mingyue Xu[1], Jian Li[5], Wenting Han[2], Chao Yang[6], Fang Li[7], Dexun Chen[8]

[1]School of Earth and Space Sciences, University of Science and Technology of China, Hefei 230026, China

[2]School of Computer Science and Technology, University of Science and Technology of China, Hefei 230026, China

[3]CAS Center for Excellence in Comparative Planetology, University of Science and Technology of China, Hefei, China

[4]Frontiers Science Center for Planetary Exploration and Emerging Technologies, University of Science and Technology of China, Hefei, China

[5]Chinese Academy of Meteorological Sciences, Beijing 100081, China

[6]School of Mathematical Sciences, Peking University, Beijing 100871, China

[7]Jiangnan Institute of Computing Technology, Wuxi, Jiangsu, China

[8]Tsinghua University, Beijing, China

**Contents of this file**

Table S1 and Figures S1 to S4

**Table S1.** Time of each PIO_initDecomp for PIO and F-PIO.

| processes<br>I/O module | 30,000 | 60,000 | 12,000 |
|---|---|---|---|
| iAMAS with PIO | 76.6s | 310.7s | 576.6s |
| iAMAS with F-PIO | 0.11s | 0.054s | 0.027s |

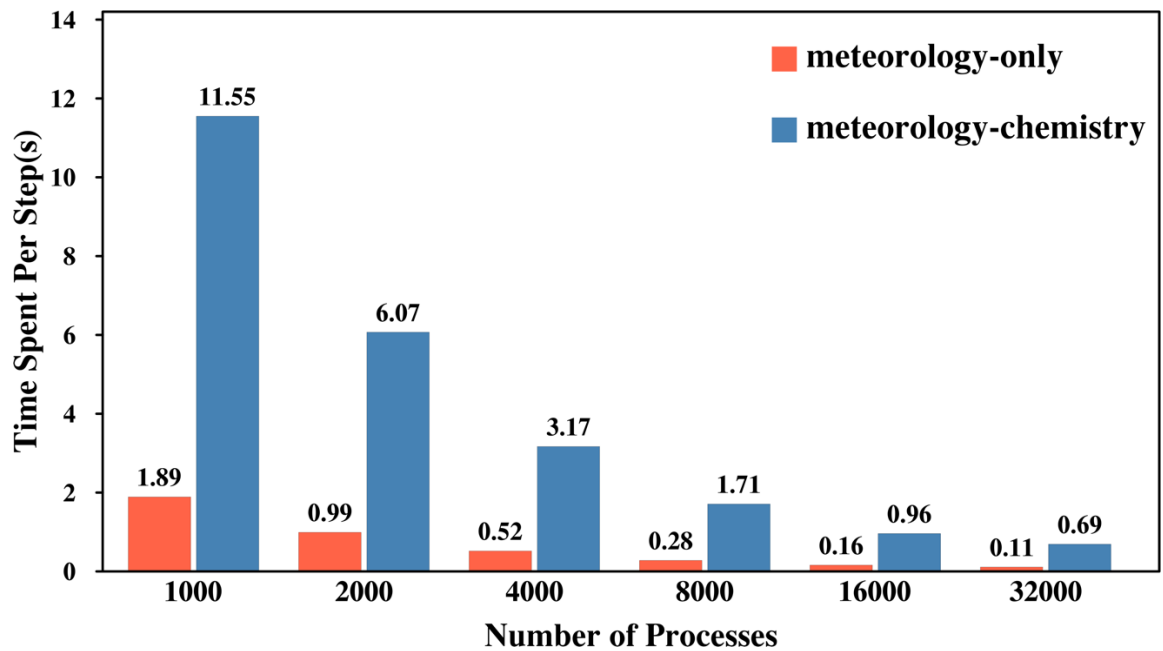

**Figure S1.** The comparison of general time required for meteorology-only and meteorology-chemistry coupled simulations.

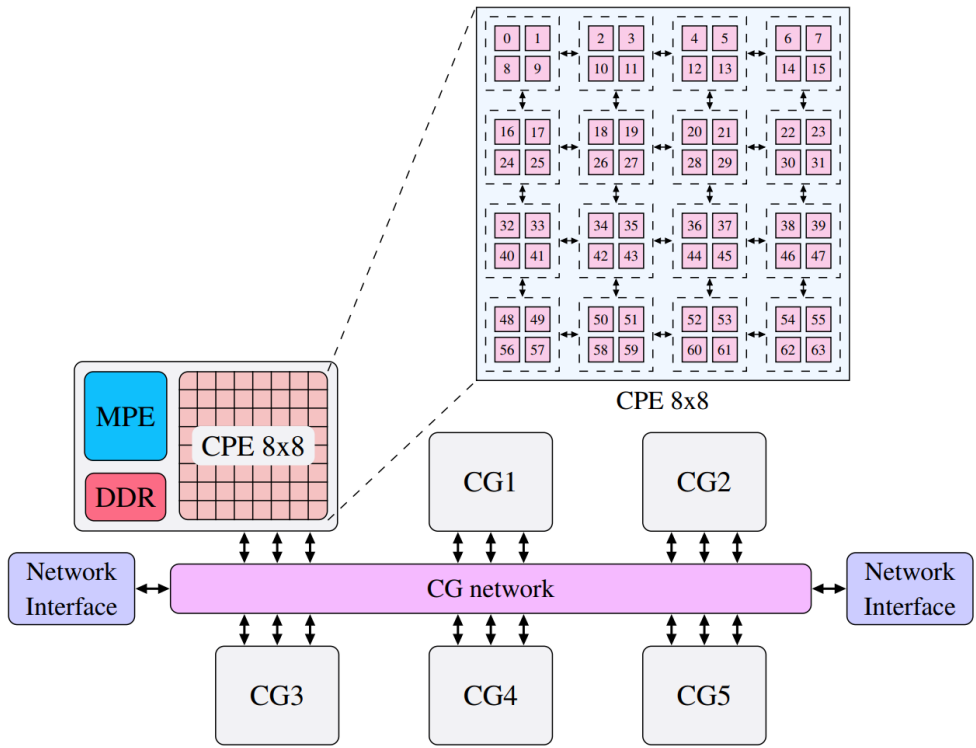

**Figure S2.** The architecture of SW26010 Pro many-core processor in the new Sunway supercomputer.

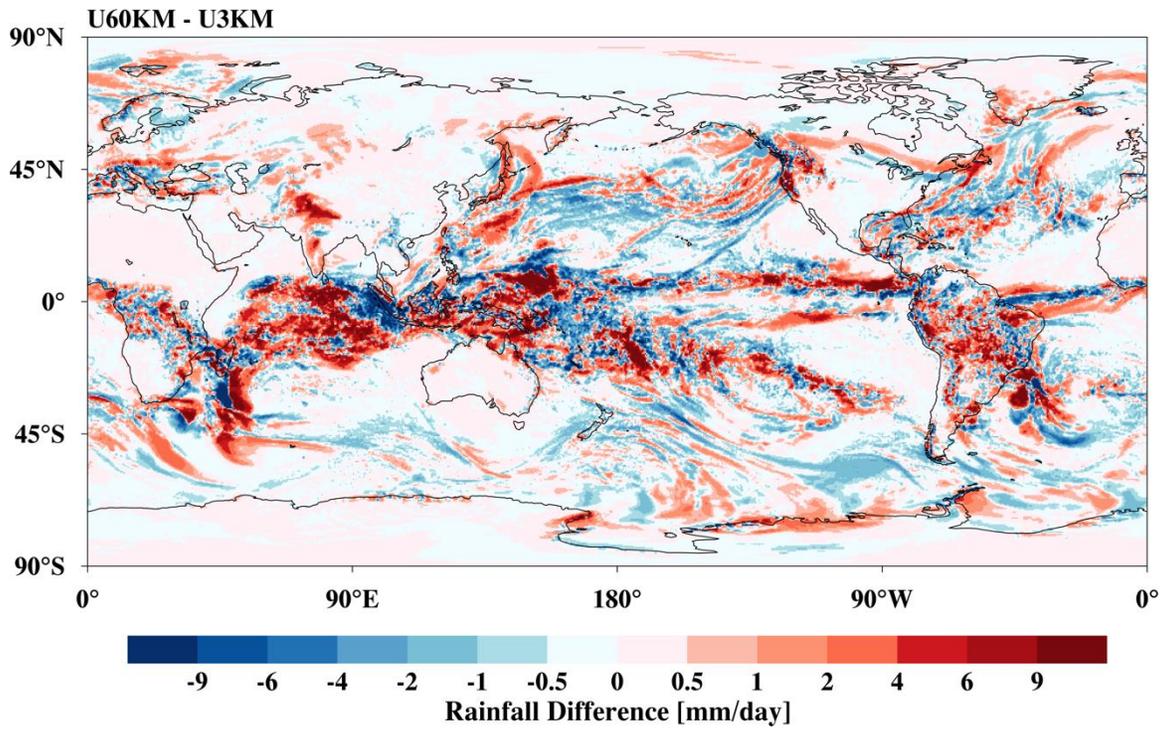

**Figure S3.** The difference between U3KM and U60KM, which is interpolated into the horizontal resolution of 0.5°.

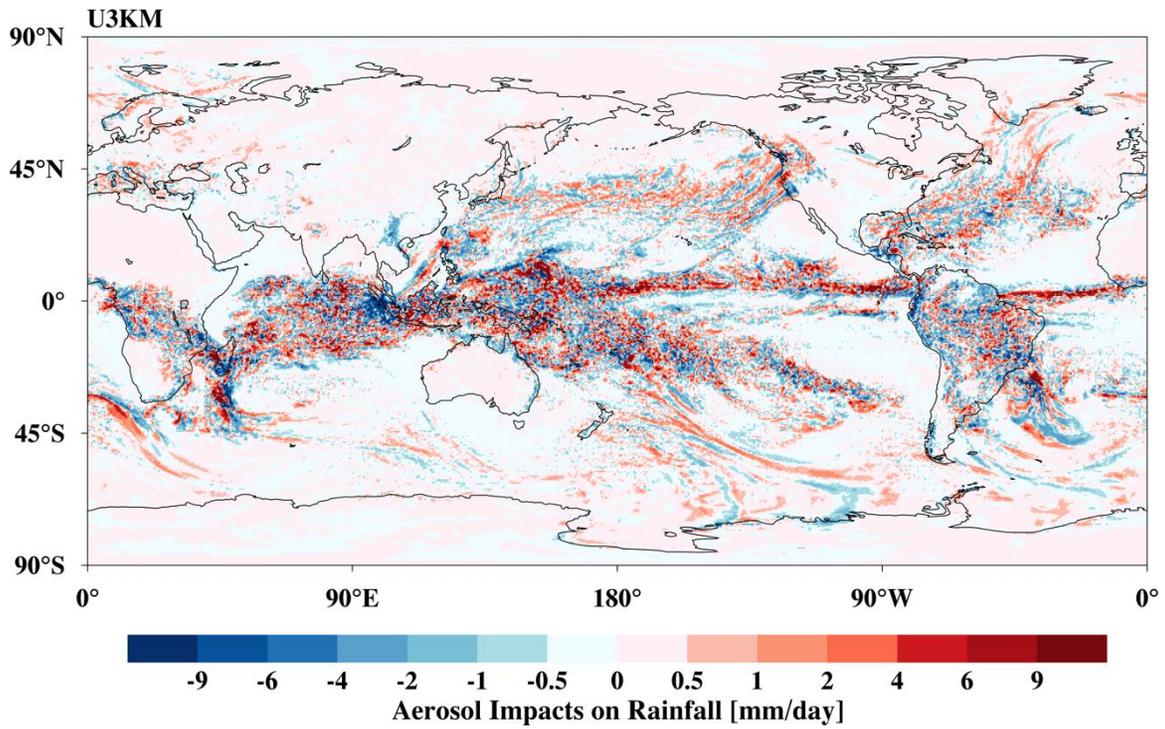

**Figure S4.** The global distributions of natural aerosol impacts on precipitation averaged during the period (Feb. 4-8) from the U3KM experiments. The results are interpolated into the horizontal resolution of 0.5°.